\documentclass[]{fairmeta}
\usepackage{graphicx}
\usepackage{amsmath}
\usepackage[version=4]{mhchem}
\usepackage{booktabs}
\usepackage{caption}
\usepackage{pifont}
\usepackage{listings}
\usepackage{adjustbox}
\usepackage{changepage}

\lstdefinelanguage{ini}{
    basicstyle=\ttfamily\small,
    morecomment=[l]{;},
    morecomment=[l]{\#},
    morestring=[b]",
    moredelim=[s][\color{blue}]{[}{]},
    alsoletter={=},
}

\tikzstyle{startstop} = [rectangle, rounded corners, 
minimum width=3cm, 
minimum height=1cm,
text centered, 
draw=black, 
fill=red!30]

\tikzstyle{io} = [trapezium, 
trapezium stretches=true, 
trapezium left angle=70, 
trapezium right angle=110, 
minimum width=3cm, 
minimum height=1cm, text centered, 
draw=black, fill=blue!30]

\tikzstyle{process} = [rectangle, 
minimum width=3cm, 
minimum height=1cm, 
text centered, 
text width=3cm, 
draw=black, 
fill=orange!30]

\tikzstyle{decision} = [diamond, 
minimum width=3cm, 
minimum height=1cm, 
text centered, 
draw=black, 
fill=green!30]
\tikzstyle{arrow} = [thick,->,>=stealth]

\usepackage[utf8]{inputenc} 
\usepackage[T1]{fontenc}    
\usepackage{hyperref}       
\usepackage{url}            
\usepackage{booktabs}       
\usepackage{amsfonts}       
\usepackage{nicefrac}       
\usepackage{microtype}      
\usepackage{xifthen}
\usepackage{multirow}
\usepackage{mciteplus}
\usepackage{epstopdf}
\usepackage{shellesc}

\newcommand{\modelcell}[3]{\parbox{2.8cm}{#1~#3}}
\usepackage{siunitx}
\usepackage{mhchem}
\usepackage{threeparttable}

\newcommand{\omc}{OMC25}
\newcommand{\mlip}{MLIP}
\newcommand{\longmlip}{machine learning interatomic potential}
\newcommand{\csd}{CSD}
\newcommand{\longcsd}{Cambridge Structural Database}

\newcommand{\sch}{Schr\"odinger}
\newcommand{\supp}{Supplementary Information}

\newcommand{\kindatiny}{\fontsize{6pt}{7.2pt}\selectfont}
\newcommand{\verytiny}{\fontsize{5pt}{6pt}\selectfont}

\newlength\savewidth\newcommand\shline{\noalign{\global\savewidth\arrayrulewidth
  \global\arrayrulewidth 0.5pt}\hline\noalign{\global\arrayrulewidth\savewidth}}
\newcommand{\tablestyle}[2]{%
    \fontfamily{ptm}\selectfont%
    \let\itold\it%
    \def\it{\itold \fontfamily{ptm}\selectfont}%
    \setlength{\tabcolsep}{#1}\renewcommand{\arraystretch}{#2}\centering\kindatiny%
    \let\citeold\cite%
    \renewcommand{\cite}[1]{\normalfont\fontfamily{ptm}\selectfont\tiny\citeold{##1}}%
}

\newcolumntype{x}[1]{>{\centering\arraybackslash}p{#1pt}}
\newcolumntype{y}[1]{>{\raggedright\arraybackslash}p{#1pt}}
\newcolumntype{z}[1]{>{\raggedleft\arraybackslash}p{#1pt}}
\newcolumntype{w}{>{\centering\arraybackslash}p{18pt}}
\newcolumntype{a}{>{\centering\arraybackslash}p{16pt}}

\definecolor{c0-title-bkg}{HTML}{ffffff}
\definecolor{c0-title-text}{HTML}{000000}
\definecolor{c0-item-bkg}{HTML}{ffffff}
\definecolor{c0-item-text}{HTML}{818589}

\definecolor{c1-title-bkg}{HTML}{d1e2dd}
\definecolor{c1-title-text}{HTML}{005953}
\definecolor{c1-item-bkg}{HTML}{e6efec}
\definecolor{c1-item-text}{HTML}{2d7b6d}

\definecolor{c2-title-bkg}{HTML}{cfe1e1}
\definecolor{c2-title-text}{HTML}{005760}
\definecolor{c2-item-bkg}{HTML}{e4eeed}
\definecolor{c2-item-text}{HTML}{24797b}

\definecolor{c4-title-bkg}{HTML}{cedce8}
\definecolor{c4-title-text}{HTML}{124e74}
\definecolor{c4-item-bkg}{HTML}{e1eaf1}
\definecolor{c4-item-text}{HTML}{3a7190}

\definecolor{c5-title-bkg}{HTML}{d0d9eb}
\definecolor{c5-title-text}{HTML}{324779}
\definecolor{c5-item-bkg}{HTML}{e1e8f3}
\definecolor{c5-item-text}{HTML}{4d6b97}

\definecolor{c6-title-bkg}{HTML}{d3d5ed}
\definecolor{c6-title-text}{HTML}{493e7b}
\definecolor{c6-item-bkg}{HTML}{e3e5f5}
\definecolor{c6-item-text}{HTML}{61639b}

\definecolor{c7-title-bkg}{HTML}{dad1ed}
\definecolor{c7-title-text}{HTML}{5a3477}
\definecolor{c7-item-bkg}{HTML}{e5e1f5}
\definecolor{c7-item-text}{HTML}{725b99}

\definecolor{c8-title-bkg}{HTML}{ded1ec}
\definecolor{c8-title-text}{HTML}{633273}
\definecolor{c8-item-bkg}{HTML}{ebe2f6}
\definecolor{c8-item-text}{HTML}{7c5997}

\definecolor{c9-title-bkg}{HTML}{e5d1eb}
\definecolor{c9-title-text}{HTML}{6c2f6b}
\definecolor{c9-item-bkg}{HTML}{f0e0f6}
\definecolor{c9-item-text}{HTML}{885591}

\definecolor{c10-title-bkg}{HTML}{ebd1e7}
\definecolor{c10-title-text}{HTML}{722e5f}
\definecolor{c10-item-bkg}{HTML}{f5e2f3}
\definecolor{c10-item-text}{HTML}{915487}

\definecolor{c3-title-bkg}{HTML}{cddfe5}
\definecolor{c3-title-text}{HTML}{330704}
\definecolor{c3-item-bkg}{HTML}{facac5}
\definecolor{c3-item-text}{HTML}{330704}

\definecolor{avg-title-bkg}{HTML}{f3f3f3}
\definecolor{avg-title-text}{HTML}{000000}
\definecolor{avg-item-bkg}{HTML}{f3f3f3}
\definecolor{avg-item-text}{HTML}{000000}

\newcommand{\addpadding}{%
  \rule{0pt}{\dimexpr\normalbaselineskip-0.5pt\relax}%
}

\newcommand{\ct}[2][c0]{\addpadding{\cellcolor{#1-item-bkg}\textcolor{#1-title-text}{#2}}}

\ExplSyntaxOn

\cs_generate_variant:Nn \coffin_typeset:Nnnnn {Nffff}

\NewDocumentCommand\rotbox{ O{l,H} D<>{0pt,0pt} m m}{
    \hcoffin_set:Nn \l_tmpa_coffin {#4}
    \coffin_rotate:Nn \l_tmpa_coffin {#3}
    \coffin_typeset:Nffff \l_tmpa_coffin 
        {\clist_item:nn{#1}{1}}
        {\clist_item:nn{#1}{2}}
        {\clist_item:nn{#2}{1}}
        {\clist_item:nn{#2}{2}}
}
\ExplSyntaxOff

\newlength{\ccustomlen}
\newcommand{\ccustom}[3][c0]{%
    \cellcolor{#1-item-bkg}{%
        \rotbox[l,t]{90}{%
            \parbox[t]{\ccustomlen}{%
                \ifthenelse{\isempty{#3}}{%
                    \mbox{%
                        \kindatiny\textcolor{#1-title-text}{#2}%
                    }%
                }{%
                    \kindatiny\textcolor{#1-title-text}{#2} \\%
                    \tiny{\textcolor{#1-item-text}{\it #3}}%
                }%
            }%
        }%
    }%
}

\newcommand{\cb}[3][c0]{%
    \setlength{\ccustomlen}{1.5cm}%
    \ccustom[#1]{#2}{#3}%
}

\usepackage{xspace}

\makeatother

\title{Open Molecular Crystals 2025 (OMC25) Dataset and Models}

\author[1]{Vahe Gharakhanyan}
\author[1]{Luis Barroso-Luque}
\author[2]{Yi Yang}
\author[1]{Muhammed Shuaibi}
\author[1]{Kyle Michel}
\author[1]{Daniel S. Levine}
\author[1]{Misko Dzamba}
\author[1]{Xiang Fu}
\author[1]{Meng Gao}
\author[3]{Xingyu Liu}
\author[2]{Haoran Ni}
\author[3]{Keian Noori}
\author[1]{Brandon M. Wood}
\author[1]{Matt Uyttendaele}
\author[3]{Arman Boromand}
\author[1]{C. Lawrence Zitnick}
\author[2,4,5]{Noa Marom}
\author[1]{Zachary W. Ulissi}
\author[1]{Anuroop Sriram}

\affiliation[1]{Fundamental AI Research at Meta}
\affiliation[2]{Department of Materials Science and Engineering, Carnegie Mellon University, Pittsburgh, PA, USA}
\affiliation[3]{Reality Labs Research at Meta}
\affiliation[4]{Department of Physics, Carnegie Mellon University, Pittsburgh, PA, USA}
\affiliation[5]{Department of Chemistry, Carnegie Mellon University, Pittsburgh, PA, USA}

\abstract{The development of accurate and efficient machine learning models for predicting the structure and properties of molecular crystals has been hindered by the scarcity of publicly available datasets of structures with property labels. To address this challenge, we introduce the Open Molecular Crystals 2025 (OMC25) dataset, a collection of over 27 million molecular crystal structures containing 12 elements and up to 300 atoms in the unit cell. The dataset was generated from dispersion-inclusive density functional theory (DFT) relaxation trajectories of over 230,000  randomly generated molecular crystal structures of around 50,000 organic molecules. OMC25 comprises diverse chemical compounds capable of forming different intermolecular interactions and a wide range of crystal packing motifs. We provide detailed information on the dataset's construction, composition, structure, and properties. To demonstrate the quality and use cases of OMC25, we further trained and evaluated state-of-the-art open-source machine learning interatomic potentials. By making this dataset publicly available, we aim to accelerate the development of more accurate and efficient machine learning models for molecular crystals.}

\metadata[Dataset]{\url{https://huggingface.co/facebook/OMC25}}
\metadata[Models]{\url{https://huggingface.co/facebook/OMC25} and \url{https://huggingface.co/facebook/UMA}}
\metadata[Code]{\url{https://github.com/facebookresearch/fairchem}}
\correspondence{V.G. (\email{vaheg@meta.com}), A.S. (\email{anuroops@meta.com}), Z.W.U. (\email{zulissi@meta.com}), N.M. (\email{nmarom@andrew.cmu.edu})}

\begin{document}


\maketitle


\section*{Background \& Summary}
\label{section:background}

\begin{figure*}[t!]
    \centering
    \includegraphics[trim={0.5cm 0 0 0},clip,width=\textwidth]{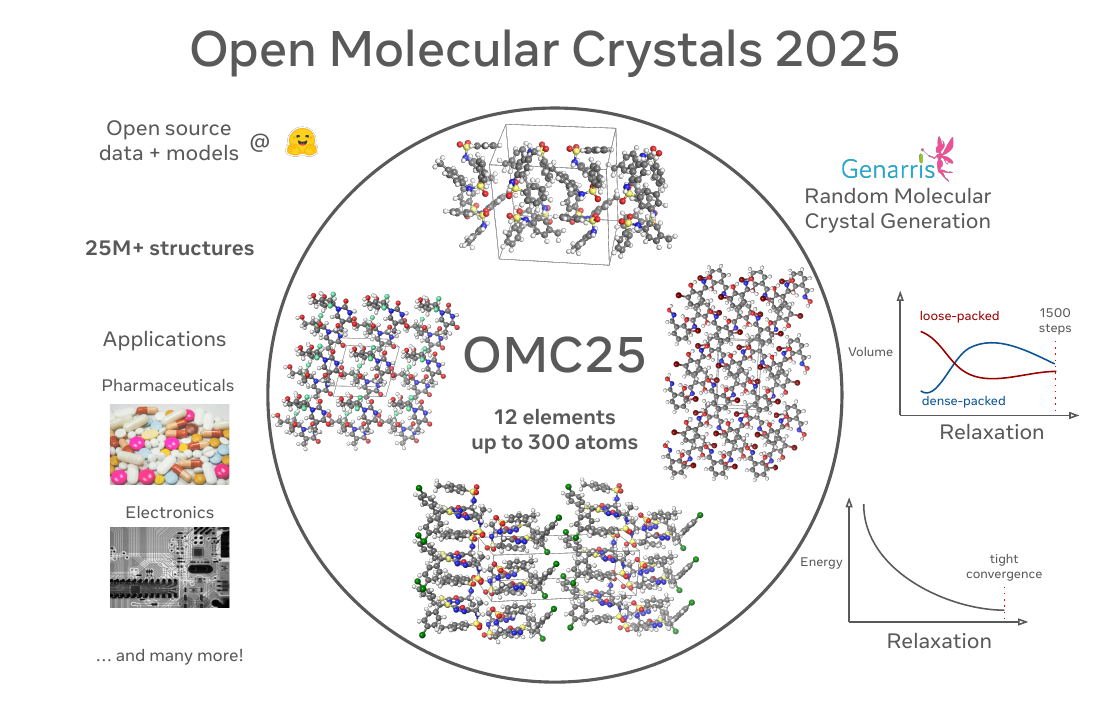}
    \caption {Overview of the OMC25 dataset: generation method, structure relaxations, statistics, and application areas.}
     \label{fig:plots}
\end{figure*}

Molecular crystals are a class of materials characterized by the orderly arrangement of molecules in a crystalline lattice. These materials have important applications in pharmaceuticals~\cite{brittain1999polymorphism, vishweshwar2006pharmaceutical, morissette2004high,rodriguez2007cocrystals,blagden2007crystal}, organic electronics~\cite{wang2018organic,dong201325th,wang2019organic,zhang2018organic}, and other fields~\cite{hao1997some,yu2019photomechanical,fang2014functional,tozawa2009porous,hunger2019industrial,niu2022crystal}, due to their unique structural and functional properties. A key phenomenon in molecular crystals is polymorphism, where a single molecule can form multiple crystal structures, influencing the physical properties of the material~\cite{lee2011crystal, bernstein2020polymorphism}.
Understanding, predicting, and controlling the formation of different polymorphs is crucial for optimizing the properties of molecular crystals for tailored applications~\cite{lee2011crystal}. This requires exploring the potential energy surfaces of different crystal structures to gain insight into their relative stability. 

Computer simulations have become indispensable tools in the study of molecular crystals. 
In recent years, the advent of machine learning (ML) has revolutionized the fields of chemistry and materials science ~\cite{wines2025chips,loew2025universal, uma,Batatia2022mace,esen,schutt2017schnet,sriram2022towards,gasteiger2021gemnet,gasteiger2022graph,schutt2021equivariant,passaro2023reducing,eqv2}. Machine learning interatomic potentials (MLIPs) trained on large \textit{ab initio} datasets offer a promising alternative to density functional theory (DFT), achieving similar accuracy at a fraction of the computational cost. The accuracy and computational demands of MLIPs typically range between those of classical force fields and DFT, depending on the model complexity and the training domain~\cite{jacobs2025practical}. This middle ground enables \mlip s to strike an effective balance between computational efficiency and accuracy, making them well-suited for large-scale simulations, such as crystal structure prediction (CSP), where using DFT would be too costly. The efficacy of ML models, however, is contingent upon the quality and diversity of the training data. The development of novel MLIP architectures has been  facilitated by the release of large, diverse, and open-source datasets tailored to molecules (OMol25~\cite{omol}, QM9~\cite{ramakrishnan2014quantum}, and others~\cite{blum2009970, axelrod2022geom, eastman2023spice, schreiner2022transition1x, smith2020ani}) and inorganic materials (OMat24~\cite{omat24}, OC20~\cite{oc20}, and others~\cite{tran2023open, odac23, kaplan2025foundational}) applications. 
MLIPs used for molecular crystals have been trained predominantly on data for isolated molecules~\cite{vzugec2024global, mann2025egret, kovacs2025mace, smith2017ani, anstine2025aimnet2}. Some have been trained on small-scale~\cite{taylor2025predictive, borysov2017organic} or proprietary~\cite{weber_efficient_2025} molecular crystal datasets. A significant gap remains in the availability of large-scale open datasets specifically designed for molecular crystal applications. This limitation hinders the advancement of MLIPs in this domain, underscoring the need for a comprehensive dataset that can provide a rich source of structural and property information for training machine learning models.

We present the Open Molecular Crystals 2025 (OMC25) dataset, a large-scale resource for training MLIPs for molecular crystals. OMC25 comprises over 27 million molecular crystal structures, containing 12 elements and up to 300 atoms in the unit cell. Each structure is labeled with total energy, atomic forces, and unit cell stress values. The structures comprising OMC25 were extracted from the dispersion-inclusive DFT relaxation trajectories of over 230 thousand putative molecular crystals constructed from 50 thousand unique molecules from the OE62 dataset~\cite{oe62}. The DFT data was acquired using the Perdew-Burke-Ernzerhof (PBE)~\cite{pbe} exchange-correlation functional combined with the Grimme D3~\cite{dftd3} dispersion correction (PBE-D3) with tight convergence settings. Diverse sampling of molecular packing arrangements across space groups with a varying number of molecules per unit cell ($Z$) was achieved using the open-source crystal generation software Genarris 3.0~\cite{genarrisv3}.  To thoroughly sample the potential energy landscape, including regions far from equilibrium, we generated  both loosely packed  and densely packed structures. To demonstrate the usefulness of the OMC25 dataset, we train MLIPs and evaluate their performance on established community benchmarks. 

The dataset, model checkpoints, and code to train and evaluate models are all released open source to ensure reproducibility and to allow the community to build upon and further improve our results. We provide the OMC25 dataset with a CC BY 4.0 license and model weights with a commercially permissive license (with some geographic and acceptable use restrictions). By making this dataset openly available, we aim to catalyze further research and enable advancements in structure and property prediction of molecular crystals.
\section*{Methods}
\label{sec:methods}

As \longmlip s (\mlip s) see increasing adoption for use in molecular crystal research, an open, comprehensive dataset specifically designed to cover molecular crystals has become an urgent need.
To achieve this, we employed a multi-step process to curate, pre-process, label, and validate a diverse set of molecular crystal structures, covering various chemical compositions, crystal systems, and space groups. The resulting \omc\ dataset includes a wide range of property-labeled molecular crystal structures, reflecting the rich diversity of molecular crystals found in nature and synthetic materials, and is designed to be a valuable resource for advancing materials research.

\textbf{Sampling Molecules.} The OE62 dataset~\cite{oe62} served as our starting point for sampling molecular structures. The OE62 dataset includes molecules extracted from the \longcsd\ (\csd)~\cite{csd} repository of experimentally determined molecular crystal structures. The OE62 dataset contains 61,489 molecules comprising the elements \ce{H}, \ce{Li}, \ce{B}, \ce{C}, \ce{N}, \ce{O}, \ce{F}, \ce{Si}, \ce{P}, \ce{S}, \ce{Cl}, \ce{As}, \ce{Se}, \ce{Br}, \ce{Te}, and \ce{I}, whose geometry was optimized with DFT through the FHI-aims all electron code~\cite{blum2009ab, ren2012resolution, zhang2013numeric} with the Perdew, Becke, and Ernzerhof (PBE) functional~\cite{pbe} and the Tkatchenko-Scheffler (TS) dispersion correction~\cite{dftts} until the residual forces on each atom were below 0.001 eV/\AA. This dataset was sampled for molecular geometries. Potentially energetic molecules were removed, resulting in approximately 50 thousand unique molecules. Owing to the scarcity of distinct conformers for a given molecule in the OE62 dataset, only one molecular conformer was obtained in the vast majority of cases. This means that, within our data generation workflow, the molecular conformation could change only to the extent possible during final geometry relaxation in the crystal, as described below.

\textbf{Molecular Crystal Structure Generation.} Random molecular crystal generation was performed with the Genarris 3.0 package~\cite{genarrisv3}. Genarris is an open-source software that generates diverse molecular crystal packing arrangements starting from the input molecular conformer and $Z$ number, the number of molecules in the unit cell. For each sampled molecule from the OE62 dataset, we selected two values of $Z$ out of the six most frequent $Z$ numbers in the \csd\ ($Z \in [4, 2, 8, 1, 16, 6]$, in order of prevalence), with the probability distributions derived from the \csd~\cite{csd}.
Genarris automatically identifies all compatible space groups matching the point group symmetry of the input molecular conformer and $Z$ number with one molecule in the asymmetric unit of the crystal ($Z^\prime$=1)~\cite{genarrisv2}. For the purpose of generating a diverse set of random structures, as opposed to performing crystal structure prediction, we did not aim to exhaustively sample the configuration space of structures within a given space group. Therefore, for each compatible space group, we generated only two structures. 

In order to train MLIPs that can work well in all regions of the potential energy surface (PES), we took advantage of the features of Genarris 3.0 to generate both loosely packed and close-packed structures. Genarris samples unit cell volumes from a Gaussian distribution around a target volume, which is estimated using the integrated machine learning (ML) model from \textsc{PyMoVE}~\cite{bier2020machine}. To generate loosely-packed initial structures, we scaled the target volume by a factor of 1.25. Genarris avoids generating unphysical structures by demanding that the interatomic distances, $d_{ij}$, between atoms $i$ and $j$ from different molecules are greater than a cutoff: $d_{ij} > s_r (r^{\text{vdW}}_i + r^{\text{vdW}}_j)$, where $r^{\text{vdW}}_{i/j}$ are the van der Waals radii of atoms $i$ and $j$, respectively, and $s_r$ is a user-defined scaling factor. In the initial generation step, we used $s_r=0.95$ to ensure adequate spacing between molecules. Then, the initial loosely packed structures were optimized with the \emph{Rigid Press} algorithm, implemented in Genarris 3.0~\cite{genarrisv3} to achieve close packing. Rigid press freezes the molecular geometry and employs a regularized hard-sphere potential to optimize the molecular position and orientation along with the crystal lattice vectors to compress the unit cell as much as possible, while preserving the space group symmetry. For the Rigid Press optimization we applied the default values of $s_r=0.85$ and specialized $s_r$ values for hydrogen bonds, derived from the statistical analysis of crystal structures in the \csd\ to ensure that molecules are as close to each other as possible without overlap~\cite{genarrisv2}.
 
\textbf{Sampling Random Molecular Crystal Structures.} To maximize the diversity of the selected putative structures, we sampled both loosely packed structures from the initial generation stage and close-packed structures from the Rigid Press stage of Genarris. The maximum number of atoms in the unit cell was capped at 300. For each compound and chosen $Z$ number, we selected up to two structures (where possible) whose geometry was converged within 5,000 iterations in the Rigid Press stage. Additionally, we sampled up to two structures from the initial generation stage, excluding the structure identifiers already sampled from the Rigid Press stage. This procedure leads to up to four structures for a given $Z$ number. As noted above, we selected two $Z$ numbers per molecule, producing a maximum of 8 putative crystal structures for each molecular conformer. In practice, only 4.7 structures on average were sampled per molecule. This approach allowed us to maximize the diversity of putative structures while minimizing redundancy, resulting in around 230,000 sampled putative crystal structures from Genarris.

\textbf{Structural Relaxations.} The molecular crystal structures selected in the previous step were fully relaxed using dispersion-corrected DFT. The calculations were performed using the Vienna Ab initio Simulation Package (VASP)~\cite{vasp1, vasp2, vasp3} with the projector augmented wave (PAW) pseudopotentials~\citep{paw1, paw2}. The PBE generalized gradient approximation (GGA)~\citep{pbe} was combined with the Grimme D3 dispersion correction~\citep{dftd3}. The atomic positions and lattice vectors were relaxed until the maximum per-atom residual forces were below 0.001 eV/\AA, or the relaxation required more than 1,500 steps. The total energy convergence tolerance was set to 0.001 meV and the plane-wave energy cut-off was 520 eV. The k-point grids were automatically set by the \textsc{pymatgen} library~\citep{pymatgen}. 
A detailed description of additional VASP input flags is provided in the Supplementary Information. The relaxation of the molecular crystal structures sampled from Genarris resulted in more than 300 million ionic steps and 1.5 billion electronic self-consistency steps.

\textbf{Relaxation Trajectory Filtering.} Regardless of the structural relaxation convergence, all trajectories were retained. Initially, we removed frames with non-negative energies, residual forces exceeding 50 eV/\AA, and stresses above 80 GPa from the trajectory. Subsequently, if the final volume or the volume immediately following the first step deviated from the initial volume by more than 33\%, the entire trajectory was discarded. However, if only a specific frame's volume differed by more than 33\% from the initial one, that frame alone was removed. This was done to make sure that the initially selected k-point density was still deemed sufficient for the ionic step. Rarely, we observed that in some cases molecular fragments broke apart or merged during relaxation. To detect this, the connectivity of each structure frame was represented as an undirected graph object in \textsc{NetworkX}~\cite{networkx} from the \textsc{StructureGraph} object of \textsc{graphs} module obtained using the \textsc{JmolNN} class from the \textsc{local\_env} module of \textsc{pymatgen.analysis}~\cite{pymatgen}. Different frames were compared with an exact graph isomorphism check as implemented in the \textsc{isomorphism} module of \textsc{networkx.algorithms}~\cite{networkx}. If the molecular connectivity after the first relaxation step differed from the initial connectivity, the entire trajectory was discarded. Otherwise, any frames with altered connectivity compared to the starting structure were removed. This was done to make sure that the structure optimization was not heading towards a non-realistic molecular crystal or leading to chemical reactions.

\textbf{Structure Sampling.} For the purpose of training MLIPs, it is important to sample different regions of the potential energy landscape, both around the local minima and far from them. Therefore, after filtering the relaxation trajectories, a sampling strategy was employed to select a representative subset of frames. The goal was to capture the most informative and diverse set of structures along the relaxation trajectories while minimizing redundancy. A subset of up to 100 structure frames was sampled from each of the remaining trajectories, with the goal of maximizing their absolute energy differences from the preceding structure frame. In addition to energy difference-based sampling, approximately 20 total structure frames were uniformly sampled between the first occurrences in the trajectory of structures with maximum per-atom residual forces of 0.1, 0.01, and 0.001 eV/\AA. The sampling strategy resulted in the final dataset including around 120 frames per molecular crystal relaxation trajectory, on average including around 10\% of all frames in the original trajectories. This led to a total sample size of 27 million structures from the relaxation trajectories. 

\begin{table}[h!]
\centering
\caption{Size, starting molecular crystal and molecule count of different OMC25 dataset splits.\label{tab:statistics}}
\resizebox{0.5\textwidth}{!}{
\tablestyle{3pt}{0.8}
\begin{tabular}{lcccc}\toprule
Split &Size &Molecular crystals &Molecules &Fraction \\\midrule
Train &24,870,226 &207,271 &44,403 &90\%\\
Val &1,386,816 &11,570 &2,467 &5\%\\
Test &1,358,143 &11,327 &2,467 &5\%\\
\cmidrule(r){1-4}
Total &27,615,185 &230,168 &49,337\\
\bottomrule
\end{tabular}}
\end{table}

\textbf{Dataset Splits.} The sampling strategy described above yielded the final structures included in the OMC25 dataset. To train MLIPs, we created the \omc~training, validation, and test splits. To prevent leakage and ensure data integrity, an allocation process was implemented, wherein all frames belonging to putative structures of the same compound were assigned to a single split exclusively. A 90/5/5 random sampling strategy was adopted, where 90\% of the data points were assigned to the training set (Train), 5\% were assigned to the validation set (Val), and 5\% to the test set (Test). The dataset sizes and compositions for each split of \omc~are presented in Table~\ref{tab:statistics}. MLIPs were trained and optimized using the training and validation subsets of the dataset, with their performance subsequently assessed on a held-out test set, as well as through additional evaluation tests and metrics.
\section*{Data Records}
\label{sec:data}

The training and validation splits of the OMC25 dataset and related files are available for download from HuggingFace at \url{https://huggingface.co/facebook/OMC25}, under the CC BY 4.0 license, after applying for the repository access on HuggingFace.

We prepared all data files in the Atomic Simulation Environment (\text{ASE}) Lightning Memory-Mapped Database (LMDB) database format. LMDB format~\cite{lmdb} is an efficient dataset type for large scale data storage designed around the concept of key-value pairs. Users can read and query the datasets using the \textsc{ASE} DB API~\cite{ase}, where each entry (value) is an \textsc{ase.atoms} object that includes information on lattice parameters, atomic positions and numbers, periodic boundary conditions, total structure energy, atomic forces, and unit cell stress. Additionally, for each entry, we also provide information on the \longcsd\ (\csd)~\cite{csd} reference code (\texttt{csd\_refcode}) corresponding to the molecule from the OE62 dataset (taken directly from~\cite{oe62}), the $Z$ value of the unit cell (\texttt{z\_value}), \texttt{genarris\_step} tag of either the generation (\emph{gener}) or the Rigid Press (\emph{press}) stage of Genarris 3.0~\cite{genarrisv3}, \texttt{xtal.id} unique crystal identifier among putative structures from Genarris step, and \texttt{sid} structure identifier consisting of the above information and also including the index of the structure frame in the filtered relaxation trajectory. We also provide detailed information on all unique initial molecular crystal structures that underwent structural relaxations in the \emph{omc25-starting-crystals.csv} file (Table~\ref{tab:data}).

\begin{table*}[h!]
\caption{Description of columns in \emph{omc25-starting-crystals.csv} describing all starting molecular crystal structures that underwent structural relaxations.\label{tab:data}}
\centering
\begin{tabular}{m{5.5cm}m{9cm}}
\shline
\ct[c3]{}{Column name} &\ct[c3]{}{Description}\\\shline
\texttt{csd\_refcode} &CSD reference code~\cite{csd} of molecule from OE62 dataset~\cite{oe62}\\\shline
\texttt{z\_value} & Number of molecules in the crystal unit cell\\\shline
\texttt{genarris\_step} &Sampled from generation (\emph{gener}) or Rigid Press (\emph{press}) step of Genarris 3.0~\cite{genarrisv3}\\\shline
\texttt{xtal.id} &Unique crystal identifier among putative structures from Genarris step\\\shline
\texttt{split} & Structure was included in the training (\emph{train}) or validation (\emph{val}) split\\\shline
\texttt{nframes} &Number of frames sampled from relaxation trajectory\\\shline
\texttt{mol.composition}, \texttt{xtal.composition} &Composition of molecule and crystal, respectively\\\shline
\texttt{mol.natoms}, \texttt{xtal.natoms} &Number of atoms in molecule and crystal unit cell, respectively\\\shline
\texttt{mol.mass}, \texttt{xtal.mass} &Molar mass in g/mol of molecule and crystal unit cell, respectively\\\shline
\texttt{xtal.spacegroup} &Crystal space group\\
\shline
\end{tabular}
\end{table*}
\section*{Technical Validation} \label{sec:technical}
A robust dataset for machine learning interatomic potentials (MLIPs) must comprehensively sample the relevant chemical, structural, and property spaces while maintaining high data quality. Here, we detail the diversity and validation of the OMC25 dataset, highlighting both its breadth, the rigorous controls applied, and the performance of MLIPs trained on it.

\textbf{Data Quality.} 
The data quality and consistency were ensured through stringent design choices, including a series of filtering and sampling steps, as detailed in the Methods section. 
We conducted tests to ensure tight numerical convergence of the VASP settings used for the PBE-D3 calculations, as described in the \supp. Unrealistic frames were removed from relaxation trajectories by filtering based on energy, force, and stress. Consistent k-point grid density was maintained by volume filtering. Filtering based on molecular connectivity eliminated frames, in which molecular bonds were broken or unrealistically formed. The sampling strategy included frames from different stages of the relaxation trajectory, capturing configurations both far from and near equilibrium to achieve thorough sampling of the potential energy landscape. This sampling strategy provides a comprehensive representation of the system's evolution during the relaxation process, while avoiding unnecessary redundancy in the regions near equilibrium. 
Notably, our dataset exhibits high consistency between the initial and final frames of the sampled trajectories. Out of 230,168 structures, only 1,516 (0.7\%) had the final frame with a different space group than that of the initial frame.

\begin{figure*}[t!]
     \centering
          \begin{subfigure}[a]{\textwidth}
         \centering
         \includegraphics[trim={0 0.4cm 0 2cm},clip,width=\textwidth]{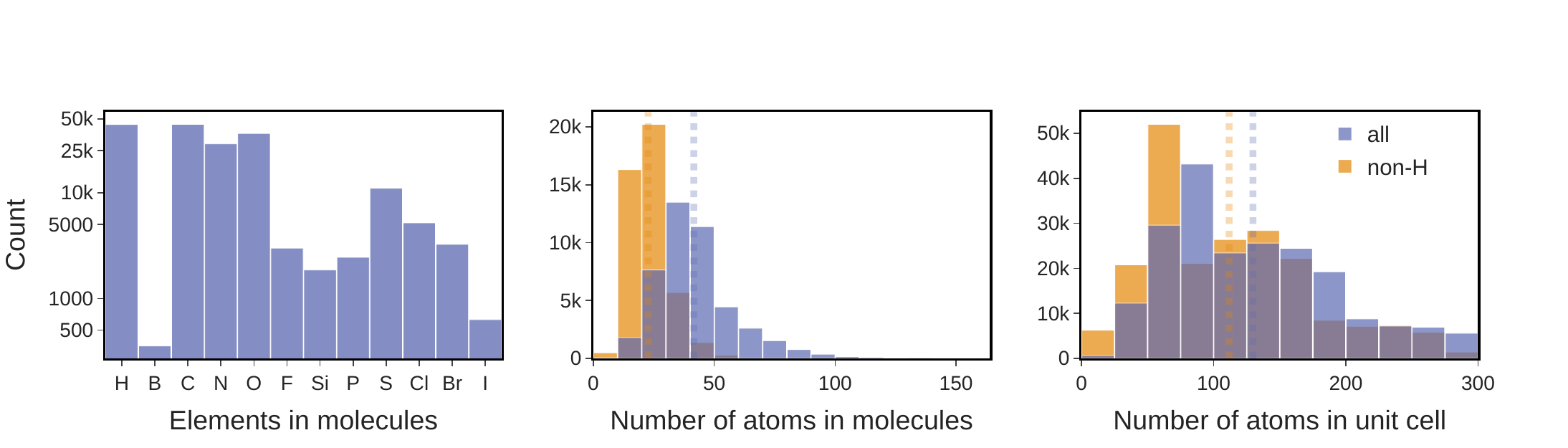}
         \caption{}
         \label{fig:dist1:a}
     \end{subfigure}
    \hfill 
     \begin{subfigure}[b]{\textwidth}
         \centering
         \includegraphics[trim={0 0 0 2cm},clip,width=\textwidth]{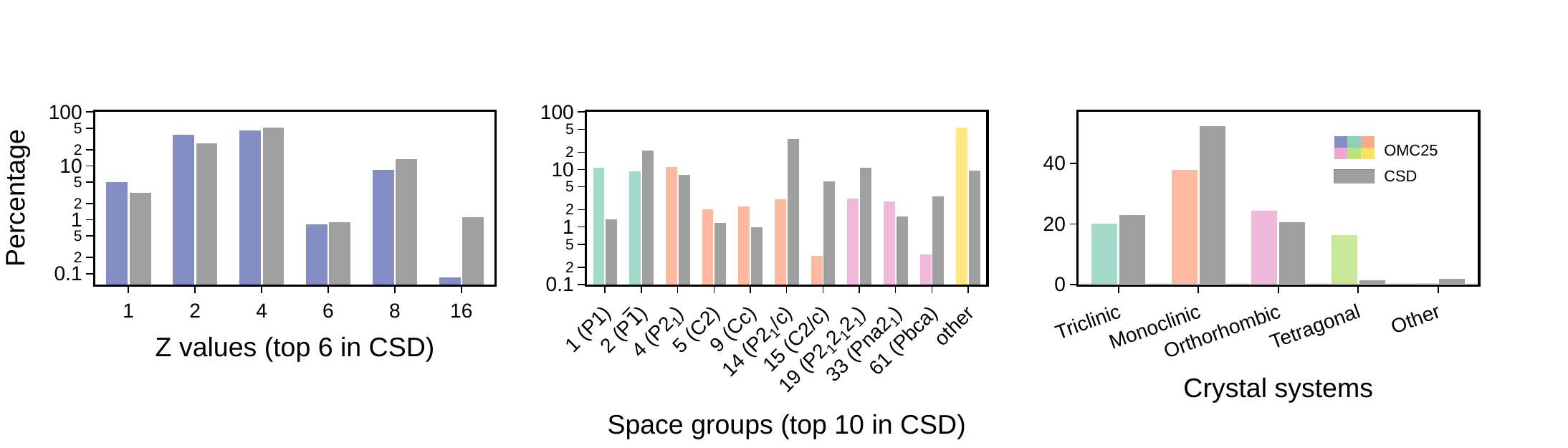}
         \caption{}
         \label{fig:dist1:c}
     \end{subfigure}
     \hfill
     \begin{subfigure}[c]{\textwidth}
         \centering
         \includegraphics[trim={2.5cm 0cm 2.5cm 1.5cm},clip,width=\textwidth]{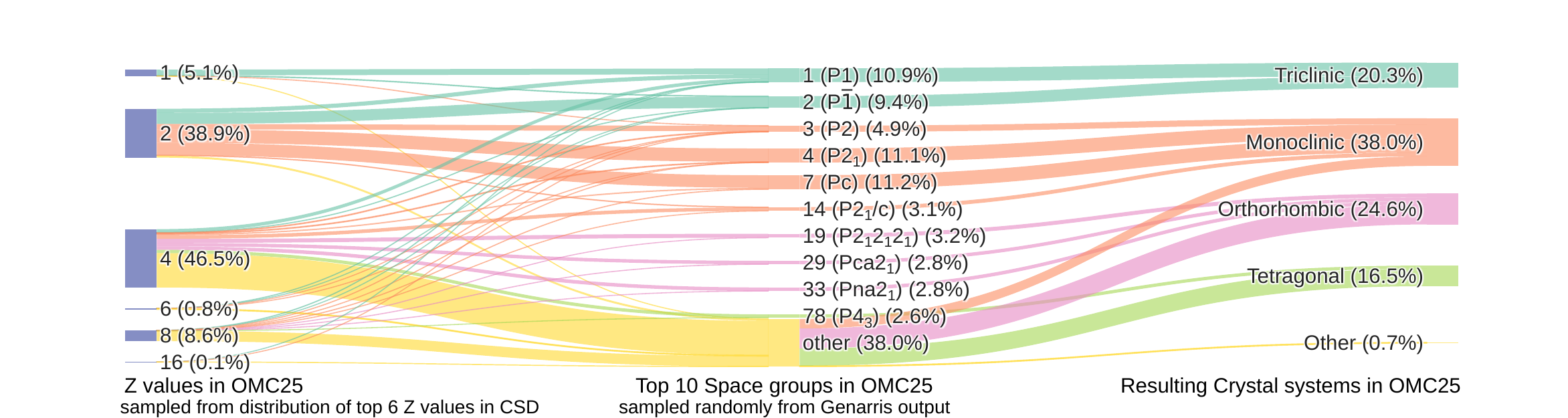}
         \caption{}
         \label{fig:dist1:d}
     \end{subfigure}
    \caption{Description of the sampled molecules and molecular crystals before structure relaxations in the OMC25 training split: \textbf{(a)} elemental occurrences in molecules, distributions (histograms) and averages (vertical dotted lines) of the number of atoms with and without counting hydrogens in starting molecules and molecular crystals, \textbf{(b)} distributions of the top six most occurring Z values, the most occurring ten space groups, and four crystal systems in organic entries of \csd\ 6.00~\cite{csd} compared to \omc, highlighted in yellow the collective distributions of the remaining space groups and crystal systems, \textbf{(c)} the Sankey diagram connecting $Z$ values, the top ten most occurring space groups in \omc, and resulting crystal systems, with the connecting links proportional to flow quantities. The coloring of space groups and connecting links is determined by crystal systems the space groups correspond to.}
     \label{fig:dist1}
\end{figure*}

\textbf{Chemical and Structural Diversity.} 
To ensure broad applicability, the \omc\ dataset was designed to capture extensive chemical and structural diversity. As shown in Figure~\ref{fig:dist1}a, the \omc\ training split encompasses 12 elements most common in organic entries of the \longcsd\ (\csd)~\cite{csd}. Figure~\ref{fig:dist1}a also shows that the number of atoms in sampled molecules ranges from 4 to 164 with an average value of 42, and the number of atoms in the crystal unit cell from Genarris ranges from 12 to 300  with an average value of 130. 

In terms of structural diversity, the \omc\ dataset includes 167 distinct space groups across all seven crystal systems. \omc\ features 22 space groups with more than 1\% of structures versus  only 10 in the CSD. In addition, there are notable differences in the prevalence of certain space groups in \omc\ compared to the CSD.  
Certain monoclinic space groups, such as $Pc$ (No. 7) and $P2$ (No. 3), are significantly overrepresented in the \omc\ dataset, accounting for 11.2\% and 4.9\% of the starting molecular crystals, respectively. In contrast, these space groups are relatively rare in the CSD, with only 0.5\% and 142 entries (out of over half a million), respectively. Similarly, the orthorhombic space group $Pcc2$ (No. 27) is represented at a frequency of 1.9\% in the \omc\ dataset, compared to only 16 entries in the CSD.
Furthermore, the tetragonal crystal system is significantly overrepresented in the \omc\ dataset, accounting for 16.5\% of the structures, whereas it is relatively rare in the CSD, with only 1.5\% representation. Conversely, the trigonal, hexagonal, and cubic systems are underrepresented in the \omc\ dataset, collectively accounting for only 0.7\% of the structures, compared to 1.9\% in the \csd. This difference may stem from  not  sampling $Z=3$ and from the lowered symmetry of the relaxed molecular structures in the OE62 dataset, which may have prevented structure generation in space groups with many special Wyckoff positions. The sampling strategy involved selecting the top six $Z$ values from the \csd, which led to a diverse set of space groups and crystal systems represented in \omc~(Figure~\ref{fig:dist1}b). The relationship between the number of molecules per unit cell and crystal symmetry is illustrated by the flow of sampled Z values into space groups and crystal systems in Figure~\ref{fig:dist1}c. This provides a visual illustration of the relationships between the number of molecules per unit cell and crystal symmetry. 

\begin{figure*}[h!]
     \centering
          \begin{subfigure}[a]{\textwidth}
         \centering
         \includegraphics[trim={0 0.3cm 0 2.5cm},clip,width=\textwidth]{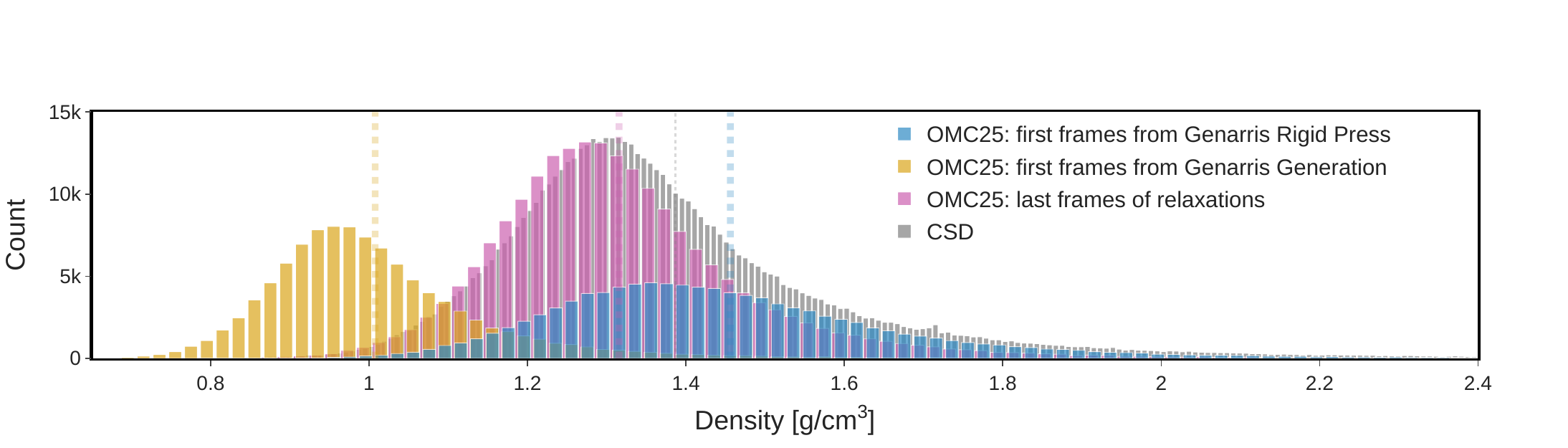}
         \caption{}
         \label{fig:dist2:a}
     \end{subfigure}
     \hfill
     \begin{subfigure}[b]{\textwidth}
         \centering
         \includegraphics[trim={0 0.25cm 0 2cm},clip,width=\textwidth]{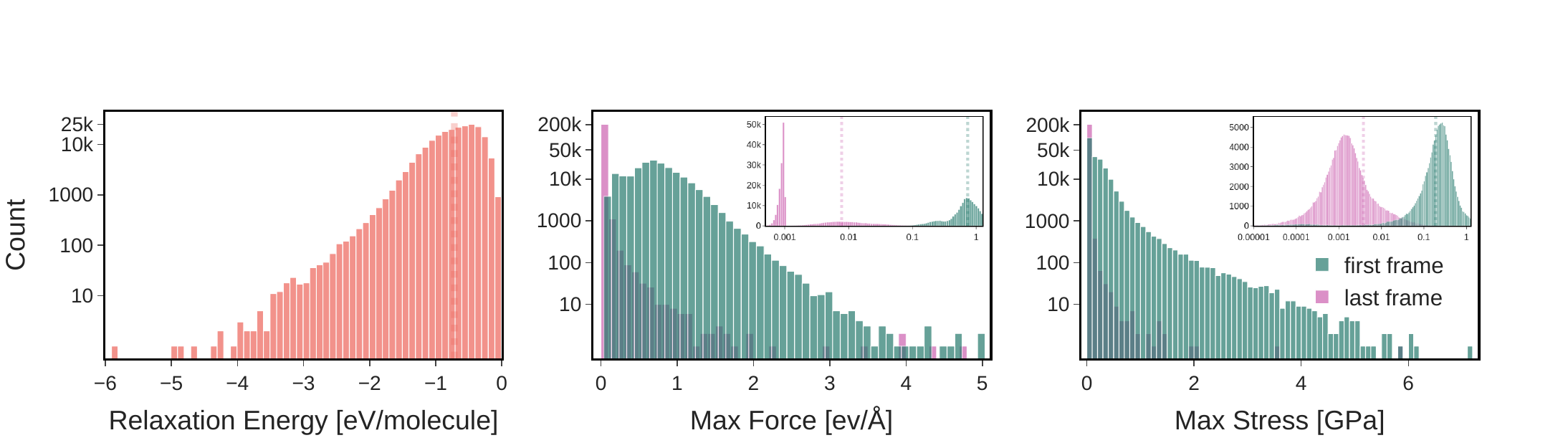}
         \caption{}
         \label{fig:dist2:b}
     \end{subfigure}
    \caption{Comparison of properties from the first and last frames of relaxation trajectories of molecular crystals in the OMC25 training split: \textbf{(a)} showing density of the last frames and the first frames for different Genarris steps; overlayed is the distribution of densities of organic entries in CSD 6.00, \textbf{(b)} showing relaxation energy (the difference of energy between the initial and final frames), maximum per-atom residual force, and maximum unit cell stress. Average values are presented with vertical dotted lines.}
     \label{fig:dist2}
\end{figure*}

\textbf{Property Diversity.} 
The analysis presented in Figure~\ref{fig:dist2} reveals extensive sampling across large swaths of the potential energy landscape, which is crucial for training robust models. 
Figure~\ref{fig:dist2}a compares the density distributions of structures at various stages: initial Genarris generation, post-Rigid Press optimization, and final relaxation. The final relaxed structures closely match the density distribution of organic CSD entries, indicating realistic packing. As expected, the structures sampled after initial generation have the lowest density. Rigid Press  effectively compresses the structures to achieve close-packing and significantly increases the density. The final relaxed structures, that closely follow CSD density distribution, are on average more dense than the as-generated structures but less dense than the structures compressed by Rigid Press. We sampled structures from both Genarris steps, ensuring that the final dataset includes almost equal number of both loosely and densely packed structures (Figure~\ref{fig:dist2}a). Starting from either loose or dense initial structures leads to different evolution of relaxation trajectories, depending on the interplay between the intensities of intramolecular and intermolecular forces. 
Figure~\ref{fig:dist2}b presents the distributions of the relaxation energy, maximum per-atom residual force, and maximum stress in the first and last frames of the relaxation trajectories that comprise the \omc\ dataset. The relaxation energy distribution highlights the extent of structural optimization, with an average difference of 0.7 eV/molecule between the initial and final structures. The force and stress distributions demonstrate that the initial structures are far from equilibrium, while the final structures show tight convergence with the VASP settings chosen, indicating mostly finished successful relaxations. These trends are consistent across training, validation, and test subsets, confirming comprehensive sampling of the potential energy landscape and underscoring the reliability and robustness of the \omc\ dataset.

\textbf{Model Performance.} 
The primary goal of the OMC25 dataset is to enable accurate, transferable predictions of molecular crystal properties using machine learning interatomic potentials (MLIPs). To this end, we adopt a model-centric validation approach: the performance of state-of-the-art ML models trained on OMC25 serves as a direct, quantitative proxy for the dataset’s informativeness, diversity, and completeness. High accuracy on relevant tasks (e.g., energy, force, and stress prediction) indicates that the dataset captures the essential physics and chemical diversity required for robust generalization. Conversely, poor model performance may reveal gaps, biases, or insufficient sampling. This approach complements traditional statistical or chemical diversity analyses and provides practical evidence of the dataset’s utility for different applications and objectives such as crystal structure prediction (CSP).

To technically validate the OMC25 dataset, we trained and benchmarked several state-of-the-art MLIPs, including UMA~\cite{uma}, eSEN~\cite{esen}, and EquiformerV2~\cite{eqv2}. These models represent the current frontier in molecular and materials modeling. Model and training parameters for eSEN and EquiformerV2 are presented in Table~\ref{tab:hyperparamaters} and UMA model details can be found in~\cite{uma}. All MLIPs are message-passing graph neural networks (GNNs) that operate on atomic graphs, with nodes representing atoms and edges representing neighboring atom pairs within a cutoff distance. We evaluate both energy-conserving models, which compute forces via automatic differentiation of the predicted energy (using PyTorch autograd~\cite{paszke2017automatic}), and direct-force models, which predict forces as an explicit output. This distinction allows us to assess the impact of model architecture on predictive accuracy.

\emph{eSEN}~\cite{esen}: We selected the eSEN energy-conserving model due to its top ranking on the Matbench Discovery leaderboard~\cite{matbench} for inorganic materials and its strong performance for the Open Molecules 2025 (OMol25) dataset~\cite{omol}. Given its success across both inorganic materials and molecular systems, we expect it to perform well for molecular crystals.

\emph{Universal Model for Atoms (UMA})~\cite{uma}: UMA is a versatile model trained on diverse datasets and atomic system types, including the \omc\ dataset. Its architecture integrates the strengths of eSEN with a mixture of linear experts, enhancing adaptability and accuracy. We focus on the OMC task within UMA and evaluate energy-conserving UMA models of small and medium sizes to find an optimal balance between performance and complexity.

\emph{EquiformerV2}~\cite{eqv2}: This direct-force GNN model incorporates transformer-inspired attention mechanisms and currently leads the Matbench Discovery leaderboard among direct-force models. We train EquiformerV2 using both a standard 6 Å cutoff and an extended 12 Å cutoff to compare the effects of model type (direct vs. energy-conserving) and receptive field size (i.e., the chemical environment an atom interacts with) on performance.

\begin{table*}[b!]
\centering
\caption{MLIP evaluations: validation and test metrics, as well as X23b and \sch\ polymorph ranking evaluations. The energy, force, and stress errors are presented as mean absolute errors (MAEs) for the validation and test metrics. For the X23b evaluation, the energy represents the lattice energy of the crystal, and the mean absolute percentage error (MAPE) is reported for the molar volume. For the \sch\ polymorph ranking evaluation, the energies are normalized to the number of molecules in the unit cell. The bolded values show the best performing models, where the evaluation followed the community standards.}
\vspace{10pt}
\label{tab:models}
\resizebox{1\textwidth}{!}{
\tablestyle{1pt}{1.05}

\begin{tabular}{y{65}wwwwwwwwwwwwwww}
\shline

\multirow{3}{*}{\vspace{-3.3cm}Model}
& \multirow{3}{*}{\vspace{-2.0cm}\cb{Number of Parameters}{}}
&\multirow{3}{*}{\vspace{-2.0cm}\cb{Conserving model}{}}
& \multicolumn{3}{c}{\ct[c4]{\it Validation}}
& \multicolumn{3}{c}{\ct[c5]{\it Test}} 
& \multicolumn{2}{c}{\ct[c6]{\it X23b~\cite{x23b}}} 
& \multicolumn{3}{c}{\ct[c7]{\it \sch}} 
\\

&
&
& \multicolumn{3}{c}{\ct[c4]{}} 
& \multicolumn{3}{c}{\ct[c5]{}} 
& \multicolumn{2}{c}{\ct[c6]{}} 
& \multicolumn{3}{c}{\ct[c7]{\it polymorph ranking~\cite{zhou2025robust}}} 
\\

&
&
& \cb[c4]{Energy $\downarrow$}{[meV/atom]} & \cb[c4]{Forces $\downarrow$}{[meV/\AA]} & \cb[c4]{Stress $\downarrow$}{[meV/\AA$^3$]} 
& \cb[c5]{Energy $\downarrow$}{[meV/atom]} & \cb[c5]{Forces $\downarrow$}{[meV/\AA]} 
& \cb[c5]{Stress $\downarrow$}{[meV/\AA$^3$]} 
& \cb[c6]{Lattice Energy}{MAE [kcal/mol] $\downarrow$} 
& \cb[c6]{Volume}{MAPE [\%] $\downarrow$}
& \cb[c7]{Rel. Energy}{MAE [kcal/mol] $\downarrow$} 
& \cb[c7]{}{Correlation \\ Pearson $\uparrow$} 
& \cb[c7]{}{Rank corelation \\ Spearman $\uparrow$}
\\

\hline
\modelcell{UMA-S-1.1 (OMC)}{}{~\cite{uma}} & 6M{\verytiny$^\dagger$} & \ding{51} & 1.05  & 5.18 & 0.95  & 1.03	& 5.04 & 0.93 &2.21 &6.01 &\textbf{0.35}   &\textbf{0.80} &\textbf{0.74} \\

\modelcell{UMA-M-1.1 (OMC)}{}{~\cite{uma}} & 50M{\verytiny$^\dagger$} & \ding{51} & 0.86  &\textbf{2.92} & 0.92 &  0.84 & \textbf{2.83} & 0.90&\textbf{1.94} &5.78 & 0.44& 0.73 &0.68\\

\modelcell{eSEN-S-OMC}{}{~\cite{esen}} & 6M & \ding{51} & 1.06	& 5.58 & 0.96 & 1.05	& 5.39 & 0.94 &3.38 &5.58 &1.04 &0.76 &0.72\\

\modelcell{eqV2-S-OMC (6 \AA)$^\ddagger$}{}{~\cite{eqv2}} & 31M & \ding{55} &\textbf{0.61}  & 3.89& 0.11 &0.70 & 3.87& 0.11&8.87 &4.02 &0.39 &0.78&0.74\\

\modelcell{eqV2-S-OMC (12 \AA)$^\ddagger$}{}{~\cite{eqv2}} & 31M & \ding{55} &0.62	&3.79	&\textbf{0.10}	&\textbf{0.67}&3.78&\textbf{0.10}& 9.09 &\textbf{2.50} &0.34 &0.79  &0.74\\

\shline
\end{tabular}
}
\begin{tablenotes}
\item \verytiny {{\tiny$^\dagger$} Reported is the number of active parameters during inference, which is lower than the total number of parameters used to train UMA models~\cite{uma}.}
\item \verytiny {{\tiny$^\ddagger$} For the X23b benchmark, the single point energies of starting molecular structures in the gas phase were taken as the reference for lattice energy calculations. For the \sch\ polymorph ranking, the single point energies of starting crystals were taken as the reference for relative energy calculations.}
\end{tablenotes}
\end{table*}

All MLIPs were evaluated on a held-out test subset, as well as on additional external benchmarks including the X23b benchmark~\cite{x23b} and the \sch\ polymorph ranking task~\cite{zhou2025robust} described below. The results of these evaluations are summarized in Table~\ref{tab:models}, with further details provided in the \supp.

\emph{X23b benchmark}~\cite{x23b}: We benchmark our MLIPs using the X23b dataset, a revised version of the original X23 experimental set~\cite{x23}. This dataset contains 23 small to medium-sized molecular crystals featuring a variety of intermolecular interactions such as van der Waals forces, hydrogen bonding, and mixed bonding types. The evaluation task involves predicting the unit cell volume and lattice energy of each crystal at 0 K, where lattice energy is defined as the cohesive energy of the crystal relative to the isolated (relaxed) molecule in the gas phase. Figure~\ref{fig:evals_si}a shows the reference values alongside the predictions from our MLIPs for each system.

\emph{\sch\ polymorph ranking}~\cite{zhou2025robust}: To evaluate the MLIPs performance for ranking polymorphs, we tested the models on a recent polymorph dataset from~\cite{zhou2025robust}. As noted in~\cite{mann2025egret}, it is important to recognize that our MLIPs were trained on data generated with the PBE-D3 functional~\citep{pbe,dftd3}, whereas the reference polymorph energies were computed using the r$^2$SCAN-D3  meta-GGA functional~\cite{furness2020accurate,dftd3}, which is considered more accurate. This discrepancy introduces some inherent limitations in the benchmark. For each system, we computed energy and rank correlation metrics, which were then averaged to produce the final evaluation scores. Distributions of these relative lattice energy metrics are presented in Figure~\ref{fig:evals_si}b.

Our results show that the OMC25 dataset can be used to train highly accurate machine learning models, with low energy, force, and stress MAEs, for the X23b evaluation. The trained MLIPs are also well-suited for the polymorph ranking task, even considering that the reference dataset was built at a higher level of density functional theory. It is important to note that the direct force models tend to perform poorly for relaxation tasks, and thus are not recommended for such applications.

To emphasize the critical role of crystal-specific data, we conducted a comparative evaluation of identical MLIP architectures trained on the OMC25 crystal dataset and the OMol25 molecular dataset~\cite{omol}. We note that the OMol25 dataset was acquired using the $\omega$B97M-V range-separated hybrid meta-GGA functional with the VV10 non-local treatment of dispersion interactions~\cite{mardirossian2016omegab97m, hellweg2015development, rappoport2010property}, which is significantly more accurate than PBE-D3. In addition, the OMol25 dataset was acquired using Gaussian basis sets without pseudopotentials of light atoms, although this is expected to have a more minor effect than the accuracy of the exchange-correlation functional and dispersion method. OMol25 is intended to be a general purpose molecular dataset and therefore lacks data on large clusters of organic molecules at various separations (though it does contain solvated systems and protein-ligand pockets) as those in OMC25. The differences in the performance of models trained on OMC25 vs. OMol25 may thus be attributed to the different DFT settings, the use of periodic vs. non-periodic codes, and the distributional shift of the underlying data. Our results reveal that model performance varies depending on the evaluation task (Table~\ref{tab:models_si}). For the X23b lattice energies, most UMA models achieve comparable accuracy, while the X23b unit cell volumes are best predicted with the OMol task of UMA. For the \sch~polymorph ranking task, UMA models with the OMC task show superior performance in energy metrics. This highlights that molecular and crystal datasets offer complementary advantages. This also underscores the necessity of including crystal data to fully capture the complexities relevant to molecular crystal modeling. Additional details are provided in the Supplementary Information.

\textbf{Limitations and Future Directions.} While the \omc\ dataset and the associated \mlip s offer significant advances, several limitations remain that warrant discussion and guide future research.

First, we limited \omc\ to single-component pristine organic molecular crystals with one molecule in the asymmetric unit ($Z^\prime$=1). However, other classes of molecular crystals—such as co-crystals, multi-component systems, hydrates, solvates, metal-organic frameworks, and disordered systems—are of great practical interest and involve unique intermolecular interactions that remain to be explored. Future work should also prioritize richer elemental diversity beyond that in the OE62 dataset~\cite{oe62}. Additionally, this study used a starting single geometry-optimized molecular conformer from the OE62 dataset (except for a handful of cases). Although DFT relaxes the atomic positions in a crystal, there is no guarantee that, in practice, the resulting structures will contain highly different molecular conformers. Incorporating multiple conformers in crystal generation will be an important step toward better modeling conformer interactions.

Second, the level of density functional theory (PBE-D3~\cite{pbe,dftd3}) employed for structure relaxations was deemed sufficient; nonetheless, several studies indicate that higher-level approaches—such as hybrid functionals like PBE0~\cite{pbe0} and meta-GGA functionals like r$^2$SCAN~\cite{grimme2020r2scan}—along with more advanced dispersion corrections (e.g., VV10~\cite{mardirossian2016omegab97m}, XDM~\cite{becke2005exchange}, TS~\cite{dftts}, and MBD~\cite{mbd1, mbd2})—are essential for accurately capturing crystal energetics~\cite{c21, marom2013many, x23, beran2016modeling, hermann2017first, x23b, grimme2020r2scan, o2022performance}. These enhancements are particularly important for CSP tasks, where high precision is required to distinguish polymorphs that differ by only a few kJ/mol~\cite{whittleton2017exchange, whittleton2017exchange2, hoja2018first, hoja2019reliable, price2023accurate, beran2023frontiers}.

Third, a known limitation of the current \mlip s lies in capturing long-range interactions. Message-passing layers can extend the effective receptive field beyond the cutoff distance, depending on the maximum number of neighbors allowed in the graph neural network (GNN). However, challenges persist in accurately modeling interactions in systems, such as organic-organic interfaces, with very long-ranged interactions, extending far beyond the cutoff radius. Additionally, the test and validation sets used in this study closely resemble the training data, limiting the ability to fully assess the generalizability of the \mlip s. To more rigorously evaluate these models, future benchmarks should incorporate both in-distribution and out-of-distribution test sets~\cite{omee2024structure} to better measure their transferability and robustness.

In summary, the \omc\ dataset is a unique resource, providing high-quality property labels for a rich and diverse set of molecular crystal structures. The \mlip s trained on  \omc\ demonstrate impressive accuracy in predicting molecular crystal structures and relative energies~\cite{fastcsp}. Together, they pave the way for accelerated simulations and more reliable molecular crystal structure and property predictions, opening new frontiers in molecular crystal research.
\FloatBarrier

\section*{Code Availability}
The random molecular crystal structure generation was performed using Genarris 3.0 package~\cite{genarrisv3}, which is available at \url{https://github.com/Yi5817/Genarris} and \url{https://www.noamarom.com/software/genarris} under the BSD 3-Clause license. All density functional theory (DFT) calculations were carried out with the Vienna Ab initio Simulation Package (VASP)~\cite{vasp1, vasp2, vasp3}. The scripts to generate VASP input files with OMC25 settings are publicly available at \url{https://github.com/facebookresearch/fairchem/tree/main/src/fairchem/data/omc}. The machine learning interatomic potentials trained on the OMC25 dataset are accessible from HuggingFace: eSEN model~\cite{esen} at \url{https://huggingface.co/facebook/OMC25} and UMA models~\cite{uma} at \url{https://huggingface.co/facebook/UMA}. We provide the pretrained models with a commercially permissive license (with some geographic and acceptable use restrictions). The code containing all necessary information for reproducing our training and evaluation results is publicly available at \url{https://github.com/facebookresearch/fairchem}. We encourage users to cite this paper when using the dataset or pretrained models for molecular crystals in their research.

\section*{Competing Interests}
The authors declare no competing interests.

\section*{Acknowledgments}
N.M. acknowledges support from the National Science Foundation (NSF) Designing Materials to Revolutionize and Engineer our Future (DMREF) program via award DMR-2323749. Y.Y. acknowledges support from the Frontera Computational Science Fellowship awarded by the Texas Advanced Computing Center (TACC).

\section*{Author Contributions}
V.G. generated the dataset; L.B.L. conducted the DFT convergence study; Y.Y. helped with configuring Genarris 3.0; everyone was involved in training, evaluating, and/or improving models; M.U. supervised the project in the initial stages; A.B., C.L.Z., N.M., Z.W.U., and A.S. supervised the entire project; V.G. prepared and everyone contributed to the review of the manuscript.


\bibliographystyle{assets/naturemag}
\bibliography{paper}


\clearpage
\newpage
\beginappendix

\setcounter{figure}{0}
\renewcommand{\thefigure}{S\arabic{figure}}
\setcounter{table}{0}
\renewcommand{\thetable}{S\arabic{table}}
\setcounter{lstlisting}{0}
\renewcommand{\thelstlisting}{S\arabic{lstlisting}}

\section*{Genarris Details}
\label{app:genarris}

\begin{adjustwidth}{0.5in}{0.5in}
\begin{lstlisting}[language=ini, caption={The base Genarris 3.0~\cite{genarrisv3} configuration file used for random molecular crystal generation.}, label={lst:genarris_config}]
[master]
name                        = <system CSD reference code>
molecule_path               = <path to molecular geometry file>
Z                           = <sampled Z number>
log_level                   = debug
restart                     = True

[generation] 
num_structures_per_spg      = 2 
specific_radius_proportion  = 0.95
natural_cutoff_mult         = 1.5
tol                         = 0.01
spg_distribution_type       = standard
max_attempts_per_spg        = 1000000 
unit_cell_volume_mean       = predict 
volume_mult                 = 1.25 
max_attempts_per_volume     = 100000 
generation_type             = crystal 

[symm_rigid_press] 
sr                          = 0.85
natural_cutoff_mult         = 1.2
int_scale                   = 0.1
method                      = BFGS
tol                         = 1e-3
maxiter                     = 5000
debug_flag                  = True
\end{lstlisting}
\end{adjustwidth}
\section*{DFT Details}
\label{app:dft}

VASP version 6.3~\cite{vasp1, vasp2, vasp3} was used for all calculations. Calculations were executed across various machine sizes and processor types using Elastic Compute within Meta's private cloud~\cite{Gupta_2024}, with parallelization parameters (such as \texttt{NCORE} and the number of MPI ranks) adjusted to the architecture of each server. Since these servers running on Elastic Compute can be preempted at any time, a single VASP calculation could be stopped and restarted many times before convergence criteria had been satisfied. With each restart, the POSCAR file was replaced with the CONTCAR file that was present when VASP stopped last and, as WAVECAR files were not written during these calculations, wavefunctions were re-initialized each time VASP was started. VASP inputs were generated using \textsc{RelaxSetGenerator} class from \textsc{atomate2}~\cite{ganose2025atomate2}.
For all structures, VASP 5.4 PBE pseudopotentials were selected, as they are suitable for the non-exotic elements present in this dataset. The atomic positions and lattice vectors were relaxed until the maximum per-atom residual forces fell below 0.001 eV/\AA, or the relaxation process exceeded 1,500 steps for most crystals, although a small, randomly selected subset (around 17\% of structures) was allowed to relax up to 3,000 ionic steps. The total energy convergence tolerance was set to 10$^{-6}$ eV, and the plane-wave energy cut-off was fixed at 520 eV, based on the recommended \texttt{ENCUT}=1.3$\times$\texttt{ENMAX} with maximum \texttt{ENMAX}=400 for the elements in our dataset. A maximum of 200 electronic self-consistency steps were allowed. The default k-point density in \textsc{pymatgen} (using the $\Gamma$-centered strategy) was applied. Relaxation outputs were parsed with \textsc{ASE}~\cite{ase} and validated using several simple consistency checks.\\\\

\begin{adjustwidth}{0.5in}{0.5in}
\begin{lstlisting}[mathescape,basicstyle=\small\ttfamily, caption={Example INCAR settings for the DFT relaxations with VASP.}, label={lst:incar}]
ADDGRID = True
ALGO = Normal
EDIFF = 1e-06
EDIFFG = -0.001
ENAUG = 1360
ENCUT = 520
GGA = Pe
IBRION = 2
ISIF = 3
ISMEAR = 0
ISPIN = 1
IVDW = 11
LASPH = True
LMIXTAU = True
LORBIT = 11
LREAL = False
NELM = 200
NELMDL = -10
NSW = 1500
PREC = Normal
SIGMA = 0.1
\end{lstlisting}
\end{adjustwidth}

\section*{DFT Convergence}
\begin{figure*}[htb!]
         \centering
         \includegraphics[width=0.85\textwidth]{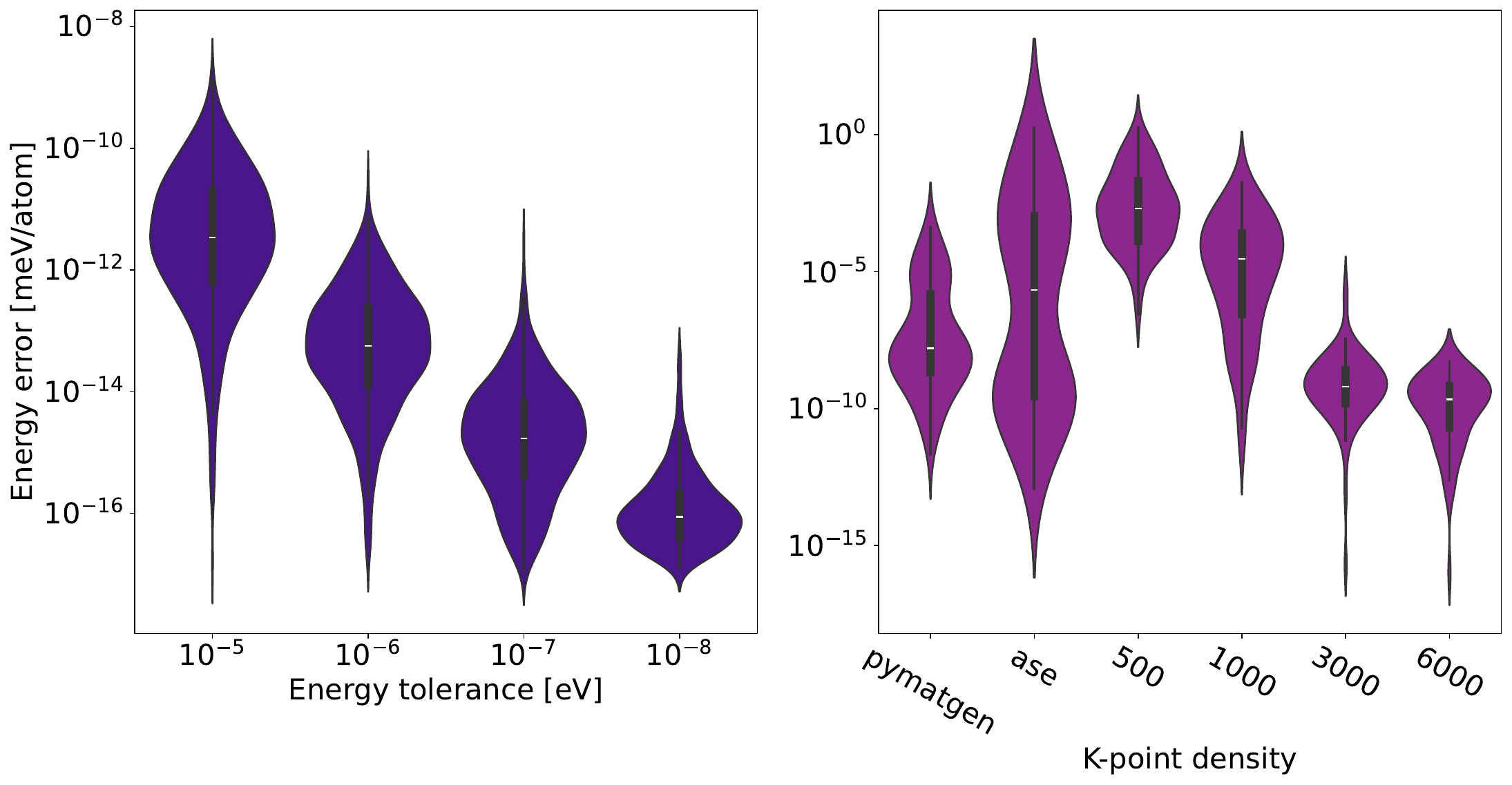}
    \caption{Convergence study for the DFT settings used in this work. We show the effect of each parameter on the energy of the structure compared to the most tightly converged settings set to 10$^{-9}$ eV energy tolerance and 9,000 k-point density per atom.}
     \label{fig:dft_convergence}
\end{figure*}

A convergence study was conducted on 500 representative molecular crystals to validate the VASP convergence criteria for total energy. Due to computational limitations, only 484 and 73 structures were used for energy tolerance and k-point density calculations, respectively. The results are presented in Figure~\ref{fig:dft_convergence}. The value used for energy tolerance (10$^{-6}$ eV) was found to be sufficiently converged for the structural relaxations reported in this work. The final energy values showed slower convergence with respect to k-point density. The error statistics for k-point meshes generated by \textsc{pymatgen} was deemed sufficient for the calculations performed. This study validates that the chosen thresholds are sufficient for the calculations.

\section*{Evaluation Details}
\label{app:eval}
\emph{X23b benchmark}~\cite{x23b}: All starting molecular crystal structures were obtained from~\cite{x23b} as low temperature polymorphs and that of ammonia and carbon dioxide crystals were taken from~\cite{ammonia} and~\cite{co2}, respectively.
We used lattice energy reference values from~\cite{x23b} but the volumes were taken from~\cite{grimme2020r2scan} to remove the effects of the revised Perdew, Burke, and Ernzerhof (RPBE)~\cite{hammer1999improved} level of theory calculations on volumes leading to unrealistically large values for selected systems. We used \textsc{ASE}~\cite{ase} and relaxed structures until the maximum per-atom residual forces were smaller than 0.001 eV/\AA, or for a maximum of 5,000 steps, constraining the relaxation to the experimental space groups. For the direct-force models, we took the single point energies of starting molecular structures in the gas phase as the reference for lattice energy calculations.

\emph{\sch\ polymorph ranking}~\cite{zhou2025robust}: For each set of polymorphs of 66 systems studied, we used \textsc{ASE}~\cite{ase} and relaxed structures with the energy-conserving models until the maximum per-atom residual forces were smaller than 0.01 eV/\AA, or for a maximum of 5,000 steps. For the direct-force models, we took the single point energies. Energy and rank correlation metrics were first computed for each system (where appropriate) and then averaged to derive the final evaluation metrics. We note that OMC25 training split contains putative structures of 24 out of 66 systems included in this benchmark.

\begin{figure*}[htb!]
     \centering
    \begin{subfigure}[a]{\textwidth}
         \centering
         \includegraphics[trim={0 0 0 0},clip,width=\textwidth]{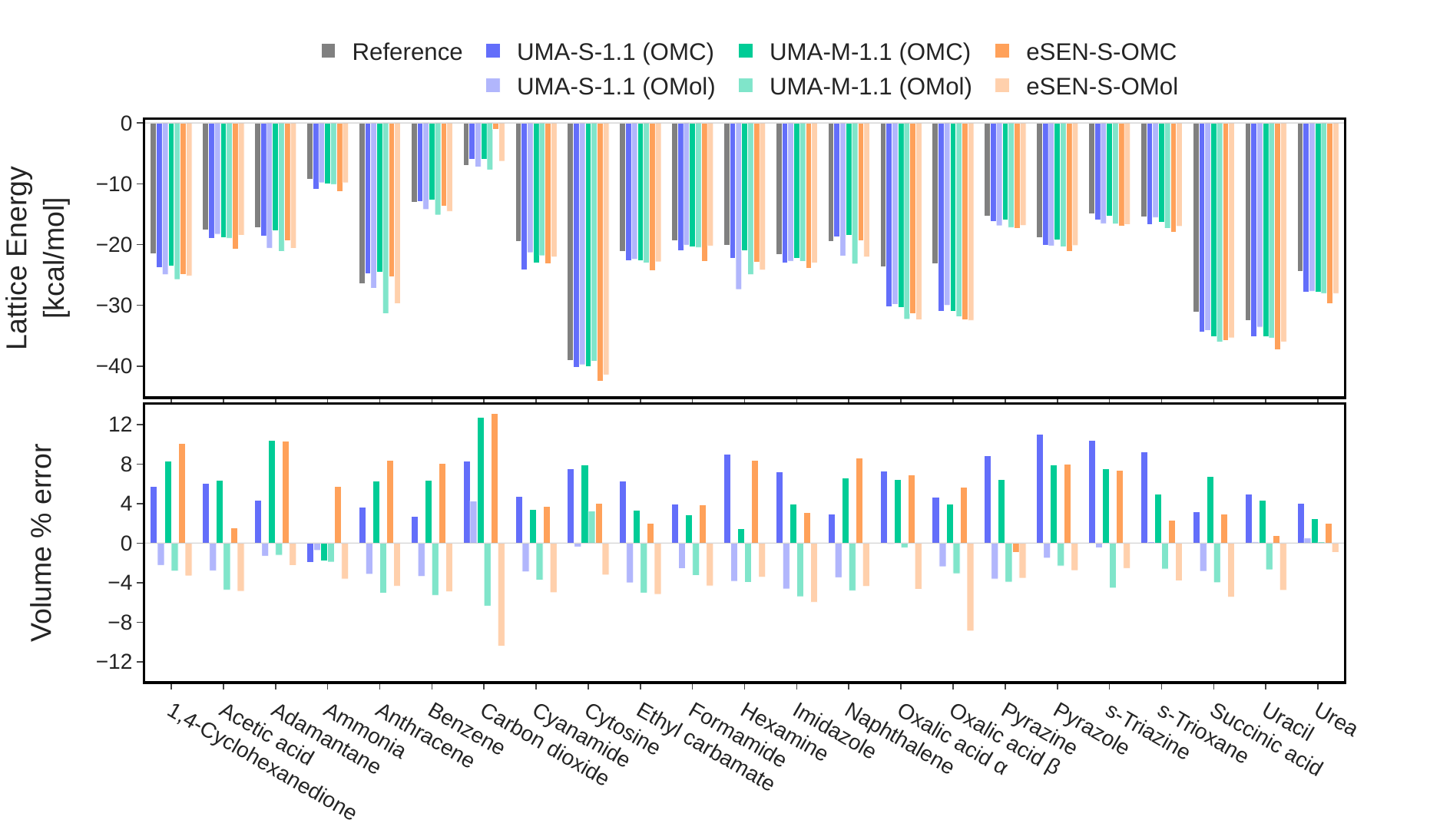}
         \caption{X23b benchmark}
         \label{fig:evals:b}
     \end{subfigure}
          \begin{subfigure}[b]{\textwidth}
         \centering
        \includegraphics[trim={1.8cm 2cm 0 0},clip,width=\textwidth]{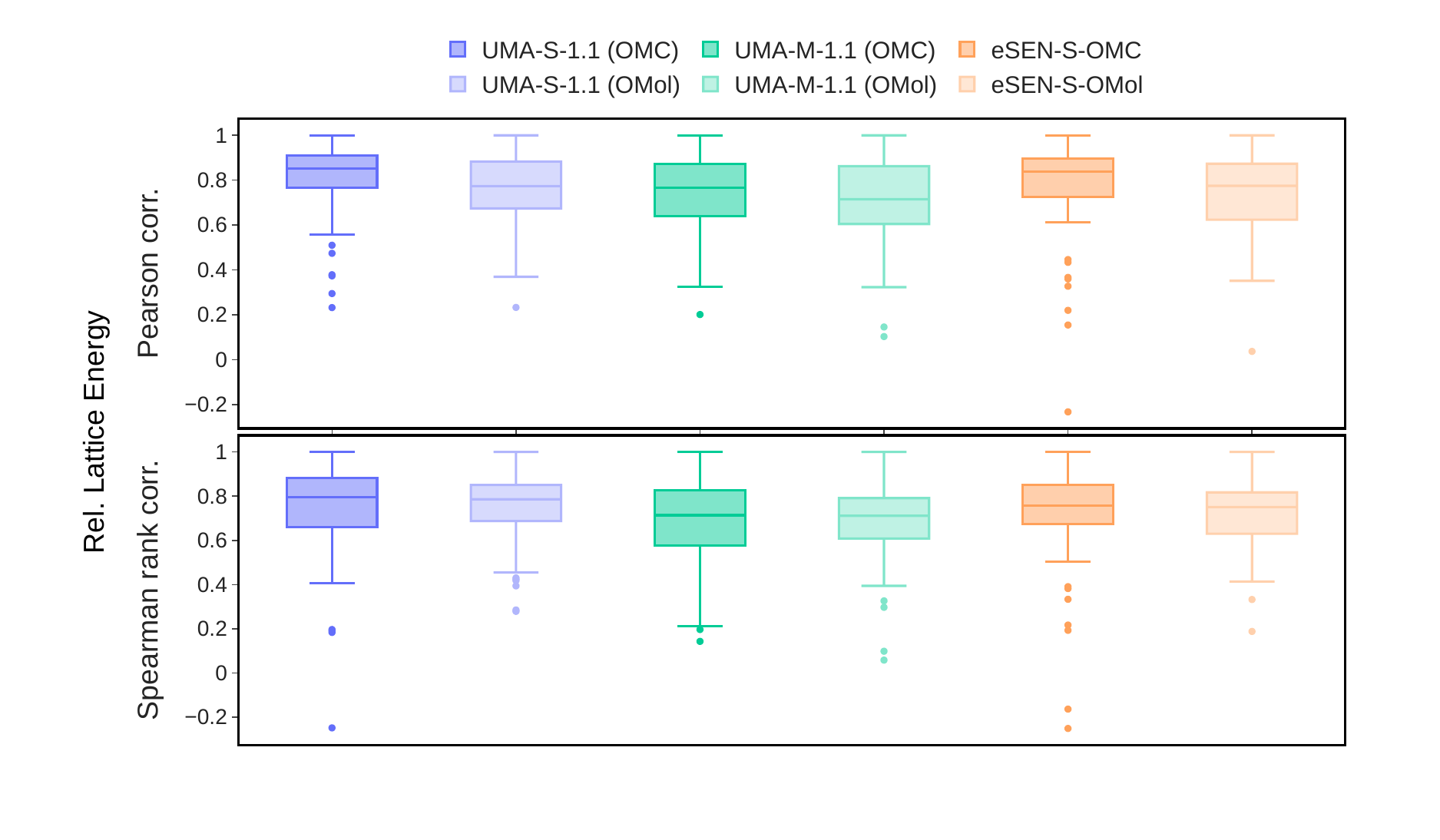}
         \caption{\sch\ polymorph ranking}
         \label{fig:evals:a}
     \end{subfigure}
    \caption{MLIP evaluation results for energy-conserving models: \textbf{(a)} per system values for the lattice energies and volume percentage errors for the X23b benchmark systems, \textbf{(b)} distributions of Pearson and Spearman rank correlations of relative lattice energies for all systems considered in \sch\ polymorph ranking.}
     \label{fig:evals_si}
\end{figure*}

\begin{table*}[h!]
\centering
\caption{Extended MLIP evaluation results for \omc\ and OMol25~\cite{omol} models: validation and test metrics, as well as X23b benchmark and \sch\ polymorph ranking task. For UMA models~\cite{uma}, we used both the OMC and OMol tasks. The bolded values show the best performing models.}
\vspace{10pt}
\label{tab:models_si}
\resizebox{1\textwidth}{!}{
\tablestyle{1pt}{1.05}

\begin{tabular}{y{65}wwwwwwwwwwy{28}y{28}www}
\shline

\multirow{3}{*}{\vspace{-3.3cm}Model}
& \multirow{3}{*}{\vspace{-2.0cm}\cb{Number of Parameters}{}}
& \multirow{3}{*}{\vspace{-2.0cm}\cb{Conserving model}{}}
& \multicolumn{3}{c}{\ct[c4]{\it Validation}}
& \multicolumn{3}{c}{\ct[c5]{\it Test}} 
& \multicolumn{2}{c}{\ct[c6]{\it X23b~\cite{x23b}}} 
& \multicolumn{5}{c}{\ct[c7]{\it \sch}} 
\\

&
&
& \multicolumn{3}{c}{\ct[c4]{}} 
& \multicolumn{3}{c}{\ct[c5]{}} 
& \multicolumn{2}{c}{\ct[c6]{}} 
& \multicolumn{5}{c}{\ct[c7]{polymorph ranking~\cite{zhou2025robust}}} 
\\

&
&
& \cb[c4]{Energy $\downarrow$}{[meV/atom]} 
& \cb[c4]{Forces $\downarrow$}{[meV/\AA]} 
& \cb[c4]{Stress $\downarrow$}{[meV/\AA$^3$]} 
& \cb[c5]{Energy $\downarrow$}{[meV/atom]} 
& \cb[c5]{Forces $\downarrow$}{[meV/\AA]} 
& \cb[c5]{Stress $\downarrow$}{[meV/\AA$^3$]} 
& \cb[c6]{Lattice Energy}{MAE [kcal/mol] $\downarrow$} 
& \cb[c6]{Volume}{MAPE [\%] $\downarrow$}
& \cb[c7]{Rel. Energy}{MAE [kcal/mol] $\downarrow$ } 
& \cb[c7]{}{RMSE [kcal/mol] $\downarrow$ }
& \cb[c7]{}{Correlation \\ Pearson $\uparrow$} 
& \cb[c7]{}{Rank corelation \\ Kendall $\uparrow$} 
& \cb[c7]{}{Rank corelation \\ Spearman $\uparrow$}
\\

\hline

\modelcell{UMA-S-1.1 (OMC)}{}{~\cite{uma}} & 6M{\verytiny$^\dagger$} & \ding{51} & 1.05  & 5.18 & 0.95  & 1.03	& 5.04 & 0.93 &2.21 &6.01 &\textbf{0.35} &\textbf{0.43}  &\textbf{0.80} &\textbf{0.60}&\textbf{0.74} 
\\

\modelcell{UMA-S-1.1 (OMol)}{}{~\cite{uma}} & 6M{\verytiny$^\dagger$} & \ding{51} & -  &- & - & - & -	& - & 2.21&\textbf{2.23} &0.55 & 0.68 &0.76 &0.59&\textbf{0.74} \\

\modelcell{UMA-M-1.1 (OMC)}{}{~\cite{uma}} & 50M{\verytiny$^\dagger$} & \ding{51} & \textbf{0.86}  &\textbf{2.92} & \textbf{0.92} &  \textbf{0.84} & \textbf{2.83} & \textbf{0.90}&\textbf{1.94} &5.78 & 0.44& 0.53&0.73 &0.55&0.68 \\

\modelcell{UMA-M-1.1 (OMol)}{}{~\cite{uma}} & 50M{\verytiny$^\dagger$} & \ding{51} & -  &- & - & - &  -  & - &3.01 &3.51 &82.6 \verytiny{(0.61$^\ddagger$)} &90.3 \verytiny{(0.76$^\ddagger$)} &0.70 &0.55 &0.69\\

\modelcell{eSEN-S-OMC}{}{~\cite{esen}} & 6M & \ding{51} & 1.06	& 5.58 & 0.96 & 1.05 & 5.39 & 0.94 &3.38 &5.58 &1.04 &1.15&0.76  &0.58 &0.72\\

\modelcell{eSEN-S-OMol}{}{~\cite{omol}} & 6M & \ding{51} & -  &- & - & - & - & - &2.85 &4.47 &0.68 &0.83  &0.73 &0.57&0.72 \\

\shline
\end{tabular}
}
\begin{tablenotes}
\item \verytiny {{\tiny$^\dagger$} Reported is the number of active parameters during inference, which is lower than the total number of parameters used to train UMA models~\cite{uma}.}
\item \verytiny {{\tiny$^\ddagger$} Reported is the result excluding five outlier systems: GLYCIN, OBEQOD, QIMKIG, QQQAUG, and UJIRIO, each with > 400 kcal/mol energy error.}
\end{tablenotes}
\end{table*}

\section*{Comparison of OMC25 and OMol25 models}
\label{app:omc_vs_omol}
To highlight the importance of crystal-specific data, we evaluated identical MLIP architectures trained on the OMC25 dataset and the molecular dataset OMol25~\cite{omol}, which contains over 100 million structures with properties computed at a higher DFT level of theory (wB97M-V/def2-TZVPD~\cite{mardirossian2016omegab97m, hellweg2015development, rappoport2010property}).

The relative performance of models trained on OMC25 and OMol25 depends on the evaluation task. For the X23b benchmark~\cite{x23b}, which involves predicting lattice energies and volumes of molecular crystals, most UMA models achieve comparable accuracy in energy predictions, with OMol25 models outperforming in volume predictions. This suggests that molecular-level data can be sufficient for predicting certain properties of crystals of small to middle-sized molecules. In contrast, for the \sch\ polymorph ranking task~\cite{zhou2025robust}, which requires precise energy ranking of closely related crystal structures, OMC25 models outperform OMol25 models in energy metrics, although their rank correlation scores are similar. These results underscore the critical role of explicit crystal environment data in capturing subtle intermolecular interactions and packing effects necessary for accurate polymorph evaluations. It is also important to note that the DFT levels of theory used for OMC25 and OMol25 differ from the reference level applied in the \sch\ polymorph ranking task. When focusing on models trained solely on a single dataset (eSEN models), OMol25-trained models perform better; however, when comparing models trained on both datasets (UMA models), those with the OMC task show superior performance. Overall, the findings demonstrate the complementary strengths of molecular and crystal datasets depending on the specific objectives.
\section*{Model Hyperparameters}
Table~\ref{tab:hyperparamaters} summarizes the model and training parameters for the eSEN~\cite{esen} and EquiformerV2~\cite{eqv2} models trained on the OMC25 dataset. The eSEN model was trained in two stages: first, a direct model with a maximum of 30 neighbors and without a stress head was trained, and, subsequently, an energy-conserving model with up to 300 neighbors and an additional stress loss was trained. Detailed descriptions of the UMA models are provided in~\cite{uma}, and the eSEN-S-OMol model is described in~\cite{omol}.

\begin{table*}[!h]
    \caption{Hyperparameters and training details for the eSEN~\cite{esen} and EquiformerV2~\cite{eqv2} models trained on the OMC25 dataset.\label{tab:hyperparamaters}}
      \centering
\scalebox{0.8}{
\begin{tabular}{l|c|c|}
\toprule
Hyperparameters & eSEN-S-OMC & eqV2-S-OMC\\ \midrule
Number of parameters & 6M & 31M\\
Maximum number of neighbors & 30, 300 & 30\\
Cutoff radius (\AA) & 6 & 6 \&\ 12\\
Number of layers & 4 & 8\\
Number of sphere channels & 128 & 128\\
Number of edge channels & 128 & 128\\
Maximum degree $L_{max}$  & 2 & 4 \\
Maximum order $M_{max}$ & 2 & 2 \\
Distance function & gaussian & gaussian \\
Number of distance basis & 64 & 512\\
Number of hidden channels & 128 &-\\
Normalization type & rms\_norm\_sh &layer\_norm\_sh \\
Activation type & gate &- \\
ff\_type & spectral &- \\
Number of Transformer blocks &- &8\\
Dimension of hidden scalar features in radial functions $d_{edge}$ &- &(0, 128)\\
Embedding dimension $d_{embed}$ &- &(4, 128)\\
$f_{ij}^{(L)}$ dimension $d_{attn\_hidden}$ &- &(4, 64)\\
Number of attention heads $h$ &- &8\\
$f_{ij}^{(0)}$ dimension $d_{attn\_alpha}$ &- &(0, 64)\\
Value dimension $d_{attn\_value}$ &- &(4, 16)\\
Hidden dimension in feed forward networks $d_{ffn}$ &- &(4, 128)\\
Grid resolution $R$ &- &18\\

\multicolumn{1}{l|}{} & \multicolumn{1}{l|}{} & \multicolumn{1}{l|}{}\\
Number of GPUs & 32 & 64\\
Optimizer & AdamW & AdamW\\
Learning rate scheduling & Cosine & Cosine\\
Warmup epochs & 0.1 & 0.01\\
Warmup factor & 0.2 & 0.2\\
Maximum learning rate & $8 \times 10 ^{-4}$ & $6 \times 10 ^{-4}$\\
Minimum learning rate factor & 0.01 & 0.01 \\
Gradient clipping norm threshold & 100 & 100\\
Model EMA decay & 0.999 & 0.999\\
Weight decay & $1 \times 10 ^{-3}$ & $1 \times 10 ^{-3}$\\
Dropout rate & - & 0.1\\
Batch size & 10,016 atoms & 76,800 systems\\
Number of epochs & 4, 2.4 & 150 \\
Stochastic depth & - & 0.1 \\
Energy loss coefficient & 10, 10 & 10\\
Force loss coefficient & 30, 2 & 5\\
Stress loss coefficient & 0, 1 & 1\\\bottomrule
\end{tabular}}
\end{table*}


\end{document}